# Strain driven anisotropic magnetoresistance in antiferromagnetic $La_{0.4}Sr_{0.6}MnO_3$

A.T. Wong,[1,2] C. Beekman,[1] H. Guo[1,3], W. Siemons,[1] Z. Gai[4], E. Arenholz[5], Y. Takamura[6], T.Z. Ward[1,3,*]

[1] Materials Science and Technology Division, Oak Ridge National Laboratory, Oak Ridge, TN, 37831, USA

[2] Materials Science & Engineering, University of Tennessee, Knoxville, TN 37996, USA

[3] Department of Physics & Astronomy, University of Tennessee, Knoxville, TN 37996, USA

[4] Center for Nanophase Materials Sciences, Oak Ridge National Laboratory, Oak Ridge, TN, 37831, USA

[5] Advanced Light Source, Lawrence Berkeley National Laboratory, Berkeley, CA, 94720 USA

[6] Department of Chemical Engineering and Materials Science, University of California, Davis, Davis, California 95616, USA

* To whom correspondence should be addressed: wardtz@ornl.gov

We investigate the effects of strain on antiferromagntic (AFM) single crystal thin films of $La_{1-x}Sr_xMnO_3$ (x = 0.6). Nominally unstrained samples have strong magnetoresistance with anisotropic magnetoresistances (AMR) of up to 8%. Compressive strain suppresses magnetoresistance but generates AMR values of up to 63%. Tensile strain presents the only case of a metal-insulator transition and demonstrates a previously unreported AMR behavior. In all three cases, we find evidence of magnetic ordering and no indication of a global ferromagnetic phase transition. These behaviors are attributed to epitaxy induced changes in orbital occupation driving different magnetic ordering types. Our findings suggest that different AFM ordering types have a profound impact on the AMR magnitude and character.

Antiferromagnets (AFM) have been shown to be a promising alternative to ferromagnets (FM) in spintronic applications.[1–5] The reason stems from the fact that at high data storage densities stray fields may destroy the FM set states while an AFM would be relatively insensitive to these stray fields and maintain its anisotropic magnetoresistance (AMR). Recent studies have focused on AFM bimetallics as they have high magnetic ordering temperatures and extremely low magnetizations.[4,5] While strongly correlated oxides are considered to have immense promise in next generation electronic and sensing devices, few studies exist for AFM spintronic applications.[5–10] In most cases, these materials have rich electron/hole doping phase diagrams populated with vastly different resistive and magnetic properties. Dopings that lie near phase boundaries are often of particular interest due to their high sensitivity to perturbations in their underlying spin-charge-orbital order parameters, which can lead to colossal changes in resistive and magnetic properties. $La_{1-x}Sr_xMnO_3$ (x = 0.6) (LSMO) is one such material. It has been shown in bulk samples to have a region of short range magnetic ordering (i.e. a deviation from Curie-Weiss behavior) between 320 K and 210 K and to straddle a canted AFM / A-type AFM phase boundary below 200 K.[11] Surprisingly, this material has not been synthesized in thin film form even though its complex phase behavior should make it highly sensitive to orbital populations controlled by epitaxy induced strain effects. While AMR in other perovskite manganites has been reported, the mechanism is typically still associated with FM phase coexistence or transition.[12–15] In this letter, we investigate strain effects on the resistive and magnetic properties of x = 0.6 doped LSMO films and find that differences in AFM ordering type can have a profound impact on the AMR magnitude and character.

Experiments were performed on 10 nm LSMO films grown epitaxially on $SrTiO_3$ (STO), $(LaAlO_3)_{0.3}$-$(Sr_2AlTaO_6)_{0.7}$ (LSAT), and $LaAlO_3$ (LAO) substrates using pulsed laser deposition.

Growth was performed in an ultrahigh vacuum system with a base pressure of < 1 x $10^{-10}$ Torr. Substrates were held at 760º C in a flowing $O_2$ environment at 0.5 mTorr. A 248 nm pulsed KrF excimer laser was operated at 1 Hz and 0.8 J/cm$^2$ fluence to give an average deposition rate of ~0.1 Å/sec. Reflection high-energy electron diffraction (RHEED) was used to monitor surfaces during growth and conformed to a layer-by-layer growth mode. After growth, samples were post annealed at 700º C in 1 atm flowing $O_2$ for 2 hours. *In-situ* RHEED and *ex-situ* atomic force microscopy confirmed flat, single phase surface morphologies. X-ray measurements were carried out on a PanAlytical X'Pert thin film diffractometer with Cu K$\alpha$ radiation. X-ray linear dichroism (XLD) measurements were conducted at beamline 4.0.2 at the Advanced Light Source in total electron yield mode by monitoring the sample drain current. The linearly polarized x-rays were incident upon the sample with a 60° angle relative to the sample normal and the *E* vector was oriented parallel or perpendicular to the scattering plane. Transport measurements were taken in a Quantum Design Physical Properties Measurement System (PPMS) using Keithley 2400 source-measure electronics at 1 µA constant current in a 4-probe geometry. Magnetization measurements were conducted on a Quantum Design Magnetic Properties Measurement System (MPMS3) in DC scan mode.

Figure 1(a) shows the *θ-2θ* XRD data containing the $002_{pc}$ LSMO and substrate peaks for each of the 3 substrates (where pc is pseudocubic). We observe that all films have grown in a uniform single phase (figure 1 inset) and present thickness fringes correlating to 10 nm thicknesses. Figure 1(b) gives reciprocal space maps taken through the $103_{pc}$ peak, which indicate that the 10 nm films are fully strained on each of the 3 substrates. In-plane a-b axes (out-of-plane c axis) lattice values are shown to be 3.906 Å (3.79982 Å ±0.00115) for STO, 3.872 Å (3.83942 Å ±0.00585) for LSAT, and 3.790 Å (3.97381 Å ±0.00769) for LAO. These dimensions are consistent with expected unit cell volumes for the x = 0.6 doping.[16] From these values, we calculate c/a ratios of 0.9728 ±0.0003 for STO, 0.9916 ±0.0015 for LSAT, and 1.0485 ±0.0020 for LAO. The offset from c/a = 1 is a known indicator of orbital state preference where modulation of the $e_g$ electron's orbital occupancy with tensile strain (c/a < 1) promotes in-plane $x^2$-$y^2$ filling, and compressive strain (c/a > 1) promotes out-of-plane $3z^2$-$r^2$ filling.[17,18] We performed x-ray linear dichroism measurements on the LSAT and LAO samples (fig. 1(c)). Integrating the area under the $L_2$ peak has been shown to be an important indicator of orbital occupancy preference.[17] Here, the compressively strained LAO sample (c/a > 1) has a positive integrated value of 0.12 indicating a $3z^2$-$r^2$ preference. The nominally lattice matched LSAT sample is very slightly tensile strained and has a negative integrated value of 0.03 which indicates a weak $x^2$-$y^2$ preference. The more strongly tensile strained STO sample is expected to continue this trend and possess a stronger $x^2$-$y^2$ preference due to its relatively smaller c/a value.

Figures 2(a)-2(c) show the resistive properties of the LSMO films from 400 K to 10 K and demonstrate a strong sensitivity to applied strain. Here, resistivity loops from 400 K→10 K→400 K were taken at 3 K/min with magnetic field along the out-of-plane direction. In each case, distinct transitions are observed; however the number and onset temperatures vary while no hysteresis is observed. Compressive strain (on LAO) induces three distinct insulating regions. Between 70 K and 400 K, little difference is observed in the insulating behavior. Below ~200K, the magnetoresistance (MR), defined as MR = [($R_0$-$R_H$)/$R_H$] x 100%, where $R_H$ and $R_0$ are respectively the resistances under field = H and 0, has a weak temperature dependence (fig. 2(d)), which then increases sharply below 70K. Since the sample resistance exceeded measurable

ranges below 60 K, the absolute MR at low temperatures may be much higher than the 40% value shown. The nominally matched LSAT sample shows four insulating regions below 400 K while presenting the strongest MR response with a maximum of 140% at 10 K. It is worth noting that unlike other manganite compositions exhibiting colossal magnetoresistance in which MR presents itself in a relatively narrow temperature region, the lattice matched LSMO film exhibits >20% MR from 240 K to 10 K.[19,20] The tensile strained film grown on STO is the only system that displays a metal-insulator transition, $T_{MIT}$ ~280 K, while a re-entrant insulator behavior occurs below 120K. We also observe that tensile strain acts to suppress MR across all temperatures with a small peak near $T_{MIT}$ and a sharp upturn below 100 K. A comparison of behaviors suggests that the unstrained film may be comprised of mixed phases, as strong MR is known to be one signature.[19] The relatively weak MR and vastly different resistive behaviors present in the tensile and compressive states at low temperatures may indicate that their preferential orbital occupations drive different but more homogeneous orderings. Since different ordering types are known to have varying sensitivity to magnetic field direction, we investigate the effects of field direction on resistivity for each of the samples.

Angular dependences of the resistivity under 9 T magnetic fields for the LSMO samples are given in figure 3. The samples were field cooled under 9 T with $\theta = 0°$ and rotation scans taken in order from the highest temperature to the lowest. Here, $\theta = 0°$ denotes the magnetic field, H, perpendicular to the current direction, J, and the sample surface, while $\theta = 90°$ corresponds to H parallel to J in the film plane. AMR can then be taken as $[(R_\theta - R_0) / R_\theta] \times 100\%$, where $R_\theta$ and $R_0$ are the resistances at $\theta$ and $\theta = 0$, respectively. Under compressive strain, there is a strong response to field direction which presents as a 2-fold symmetric curve similar to the shape and amplitude of the highest reported AMR values in ferromagnetic based manganite devices and following the same $cos^2(\theta)$ dependence.[14,21] Values range from 5% at 200 K to 63% at 60 K with maxima occurring when the magnetic field is perpendicular to the film. The sharp jump in AMR from 75 K to 60 K suggests that these percentages might be much higher at lower temperatures, but we are unable to measure due to the high resistances. The lattice matched LSAT sample also shows 2-fold symmetric behavior at high temperature, but a resistive plateau of several degrees around field perpendicular begins to present itself at 100 K that breaks the $cos^2(\theta)$ dependence shown in the compressively strained case. The AMR values range from 1% at 150 K to 8% at 10 K with the biggest jump in response occurring between 100 K and 50 K. The differences between AMR in the tensile and lattice matched samples is interesting, as the absolute MR is considerably higher in the lattice matched sample while the compressively strained sample is much more sensitive to field direction. Tensile strain in the STO sample nearly extinguishes the AMR response with values on the order of 0.1% at all temperatures below 300 K. However, at 100 K, 50 K and 10 K we observe an intriguing response characterized by sharp increases in resistance occurring near magnetic field perpendicular to film surface. The angular position of the peaks increases slowly with decreasing temperature (i.e. +/- 7° at 100K, +/- 15° at 50 K and, at +/- 17° at 10 K) while the minima occur at +/- 90° for all temperatures. This behavior is not consistent with any previous explanations involving phase coexistence or substrate miscut induced anisotropy[12,13,15,22]. Also, the relatively high 9 T magnetic field should be more than sufficient to overcome inherent magnetic anisotropy arising from non-uniform mesoscale distribution of strain caused by STO twinning effects.[23,24] STO twinning can be further ruled out as the cause by the fact that the observed effect changes with decreases temperature well below the STO phase transition. Instead, as we suggest below, these peaks may be the result of electron scattering from a canted AFM phase.

Figure 4 presents field cooled (FC) and zero field cooled (ZFC) magnetization data with H = 30 mT. The diamagnetic background from the substrate has been subtracted from the curves. In each case, an upturn in magnetization is observed below 50 K that may be caused by substrate induced paramagnetism (PM). None of the samples show a clear FM onset and all present a weak total magnetization across the entire temperature range. Without a FM transition, we can rule out PM to FM transition as the source of the MR responses which is a common avenue for strong MR and AMR in manganites[15]. The compressively strained LAO sample shows a weak signature of magnetic ordering below ~200 K corresponding to the bump in MR and the onset of strong AMR. However, there is no change in magnetic behavior corresponding to the sharp increase in resistivity below 70 K. This may signal a freezing out of the conduction along the $3z^2$-$r^2$ dominated super-exchange channel. The lattice matched LSAT sample shows a slight decrease in magnetization from 350 K to 300 K which may indicate a transition away from a high temperature PM phase similar to what is reported in bulk samples.[11] This transition also coincides with an increase in resistivity and MR in transport measurements. We observe a weak ordering signature slightly below 200 K which again coincides with a large jump in AMR. The tensile strained STO sample shows a slight increase in magnetization from 350 K to 310 K on FC that coincides with the metal-insulator transition; however magnetization vs field loops show no indication of ferromagnetism (not shown). When combined with the fact that this magnetic transition is not seen in the ZFC data and that there is a change in MR near 300 K, a region of mixed phases or short range ordering can be assumed between 310 K and 120 K.[11] At 120 K, the strongest magnetic ordering signature of the three strain states appears. It is important to note that this temperature is above the STO cubic-tetragonal phase transition at 105 K so we can conclude that the observed magnetic ordering is not the result of a substrate transition. Further, the unusual AMR responses of hard axis peaks at +/- several degrees to perpendicular magnetic field application occur below 100 K which is well within the ordered window. This is within the temperature range where a magnetically ordered canted AFM phase is known to reside in bulk, so we speculate that these resistive peaks at set angles may be the result of scattering from the ordered spin canted sites. We also note that the magnetoresistive behavior appears to conform to a spin-flop type of response for all three strain states; as neither out-of-plane ±9T magnetic field dependent resistivity loops nor ±7T magnetic field dependent magnetization loops (data not shown) present the typical sharp discontinuities that signal a spin-flip transition seen in other AFM manganites.[25,26]. Figure 5 presents low field AMR scans taken at 2 T for the compressively strained sample and 3 T for the matched and tensile strained samples at 60 K and 10 K respectively and were nearly identical to the 9 T AMR angular response curves but with lower absolute MR response magnitude which suggests that the AFM ground state has not been melted in any of the samples even at the highest field.

It is known that magnetic degrees of freedom can be indirectly controlled by tuning orbital degrees of freedom and has been shown theoretically that for the $La_{1-x}Sr_xMnO_3$ system slight variations around c/a = 1 at x = 0.6 could lead to different magnetic ground states.[27] Specifically, c/a < 1, where $x^2$-$y^2$ orbital occupation dominates, leads to an A-type AFM phase; and c/a > 1, where $3z^2$-$r^2$ orbital occupation dominates, leads to a C-type AFM phase.[28,29] This is consistent with our results where compressive strain arising from the LAO substrate has a c/a = 1.049 while the nominally lattice matched LSAT sample and tensile strained STO sample have c/a values of 0.992 and 0.973 respectively. Thus, the large differences in AMR values may be the result of these different ordering types where the coupling between orbital occupation and spin ordering dominates behavior. AMR in AFM materials has been predicted and experimentally confirmed; but only in tunnel junctions[1,3] and at much lower AMR values in simple ohmic devices[4,5] than what we observe in the compressively strained and near lattice matched x = 0.6 LSMO films. The strong field orientation results suggest that this material would be a good candidate for tunnel junction devices where spin coupling across the interface may give unprecedented response. Future studies are needed to identify the exact magnetic ordering types induced by the different strain states and whether mixed electronic phases may play a role. These studies suggest that complex oxides may find use in harnessing AMR for AFM spintronic applications.

**Acknowledgements**
This effort was supported by the US Department of Energy (DOE), Office of Basic Energy Sciences (BES), Materials Sciences and Engineering Division, (TZW, CB, and WS) and under US DOE grant DE-SC0002136 (AW, HWG). Magnetization measurements (ZG) were conducted at the Center for Nanophase Materials Sciences, which is sponsored at Oak Ridge National Laboratory by the Scientific User Facilities Division, Office of BES, US DOE. The Advanced Light Source is supported by the Direction, Office of Science, Office of BES, of the US DOE under Contract No. DE-AC02-05CH11231. YT acknowledges the support of the National Science Foundation Contract No. DMR 0747896.

**References**
[1] B.G. Park, J. Wunderlich, X. Martí, V. Holý, Y. Kurosaki, M. Yamada, H. Yamamoto, A. Nishide, J. Hayakawa, H. Takahashi, A.B. Shick, and T. Jungwirth, Nat. Mater. **10**, 347 (2011).
[2] C. Chappert, A. Fert, and F.N. Van Dau, Nat. Mater. **6**, 813 (2007).
[3] Y.Y. Wang, C. Song, B. Cui, G.Y. Wang, F. Zeng, and F. Pan, Phys. Rev. Lett. **109**, 137201 (2012).
[4] A.B. Shick, S. Khmelevskyi, O.N. Mryasov, J. Wunderlich, and T. Jungwirth, Phys. Rev. B **81**, 212409 (2010).
[5] X. Marti, I. Fina, C. Frontera, J. Liu, P. Wadley, Q. He, R.J. Paull, J.D. Clarkson, J. Kudrnovský, I. Turek, J. Kuneš, D. Yi, J.-H. Chu, C.T. Nelson, L. You, E. Arenholz, S. Salahuddin, J. Fontcuberta, T. Jungwirth, and R. Ramesh, Nat. Mater. **13**, 367 (2014).
[6] M.J. Rozenberg, I.H. Inoue, and M.J. Sánchez, Appl. Phys. Lett. **88**, 033510 (2006).
[7] K.S. Takahashi, M. Gabay, D. Jaccard, K. Shibuya, T. Ohnishi, M. Lippmaa, and J.-M. Triscone, Nature **441**, 195 (2006).
[8] H. Guo, J.H. Noh, S. Dong, P.D. Rack, Z. Gai, X. Xu, E. Dagotto, J. Shen, and T.Z. Ward, Nano Lett. **13**, 3749 (2013).
[9] L. Jiang, W.S. Choi, H. Jeen, T. Egami, and H.N. Lee, Appl. Phys. Lett. **101**, 042902 (2012).
[10] T.Z. Ward, Z. Gai, X.Y. Xu, H.W. Guo, L.F. Yin, and J. Shen, Phys. Rev. Lett. **106**, 157207 (2011).
[11] J. Hemberger, A. Krimmel, T. Kurz, H.-A. Krug von Nidda, V.Y. Ivanov, A.A. Mukhin, A.M. Balbashov, and A. Loidl, Phys. Rev. B **66**, 094410 (2002).
[12] T. Jungwirth, J. Sinova, J. Mašek, J. Kučera, and A.H. MacDonald, Rev. Mod. Phys. **78**, 809 (2006).
[13] W. Ning, Z. Qu, Y.-M. Zou, L.-S. Ling, L. Zhang, C.-Y. Xi, H.-F. Du, R.-W. Li, and Y.-H. Zhang, Appl. Phys. Lett. **98**, 212503 (2011).
[14] J.-B. Yau, X. Hong, A. Posadas, C.H. Ahn, W. Gao, E. Altman, Y. Bason, L. Klein, M. Sidorov, and Z. Krivokapic, J. Appl. Phys. **102**, 103901 (2007).

[15] R.-W. Li, H. Wang, X. Wang, X.Z. Yu, Y. Matsui, Z.-H. Cheng, B.-G. Shen, E.W. Plummer, and J. Zhang, Proc. Natl. Acad. Sci. **106**, 14224 (2009).

[16] A. Bhattacharya, X. Zhai, M. Warusawithana, J.N. Eckstein, and S.D. Bader, Appl. Phys. Lett. **90**, 222503 (2007).

[17] D. Pesquera, G. Herranz, A. Barla, E. Pellegrin, F. Bondino, E. Magnano, F. Sánchez, and J. Fontcuberta, Nat. Commun. **3**, 1189 (2012).

[18] C. Aruta, G. Ghiringhelli, A. Tebano, N.G. Boggio, N.B. Brookes, P.G. Medaglia, and G. Balestrino, Phys. Rev. B **73**, 235121 (2006).

[19] E. Dagotto, T. Hotta, and A. Moreo, Phys. Rep. **344**, 1 (2001).

[20] A.-M. Haghiri-Gosnet and J.-P. Renard, J. Phys. Appl. Phys. **36**, R127 (2003).

[21] M. Egilmez, M.M. Saber, A.I. Mansour, R. Ma, K.H. Chow, and J. Jung, Appl. Phys. Lett. **93**, 182505 (2008).

[22] M. Mathews, F.M. Postma, J.C. Lodder, R. Jansen, G. Rijnders, and D.H.A. Blank, Appl. Phys. Lett. **87**, 242507 (2005).

[23] A. Khapikov, L. Uspenskaya, I. Bdikin, Y. Mukovskii, S. Karabashev, D. Shulyaev, and A. Arsenov, Appl. Phys. Lett. **77**, 2376 (2000).

[24] A. Biswas, M. Rajeswari, R.C. Srivastava, T. Venkatesan, R.L. Greene, Q. Lu, A.L. de Lozanne, and A.J. Millis, Phys. Rev. B **63**, 184424 (2001).

[25] G. Cao, J. Zhang, Y. Xu, S. Wang, J. Yu, S. Cao, C. Jing, and X. Shen, Appl. Phys. Lett. **87**, 232501 (2005).

[26] M. Tokunaga, N. Miura, Y. Tomioka, and Y. Tokura, Phys. Rev. B **57**, 5259 (1998).

[27] Z. Fang, I.V. Solovyev, and K. Terakura, Phys. Rev. Lett. **84**, 3169 (2000).

[28] T. Akimoto, Y. Maruyama, Y. Moritomo, A. Nakamura, K. Hirota, K. Ohoyama, and M. Ohashi, Phys. Rev. B **57**, R5594 (1998).

[29] R. Kajimoto, H. Yoshizawa, H. Kawano, H. Kuwahara, Y. Tokura, K. Ohoyama, and M. Ohashi, Phys. Rev. B **60**, 9506 (1999).

**Figures**

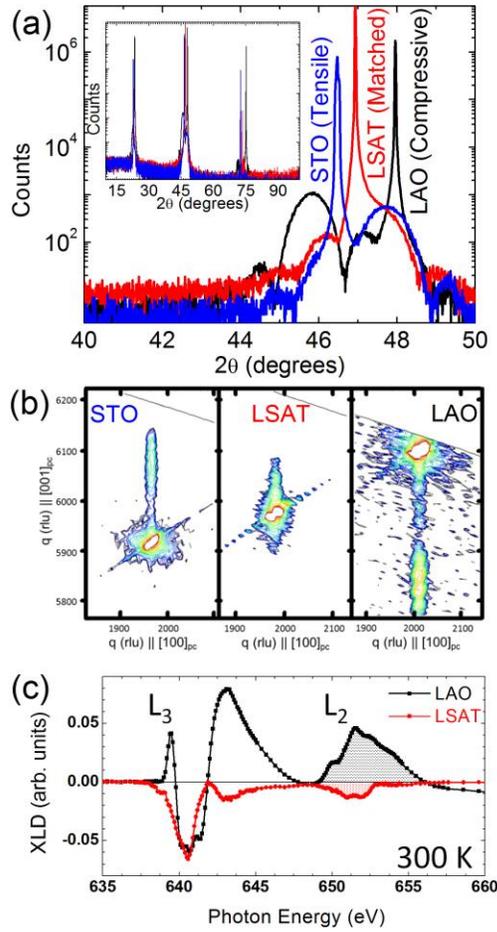

Figure 1. (a) θ-2θ XRD scans around the $002_{pc}$ peak of 10 nm thick LSMO films grown on different substrates. Inset shows longer range XRD results and present no spurious phases. (b) Reciprocal-space maps taken through the $103_{pc}$ peak to demonstrate that LSMO films are commensurate with underlying substrates. (c) XLD spectra taken at 300 K on films on LAO and LSAT substrates. Integration of $L_2$ peaks, denoted by thatched regions, show an orbital preference dependent on c/a values where compressive strain gives preference to $3z^2$-$r^2$ and weak tensile strain gives a slight preference for $x^2$-$y^2$ orbital occupation.

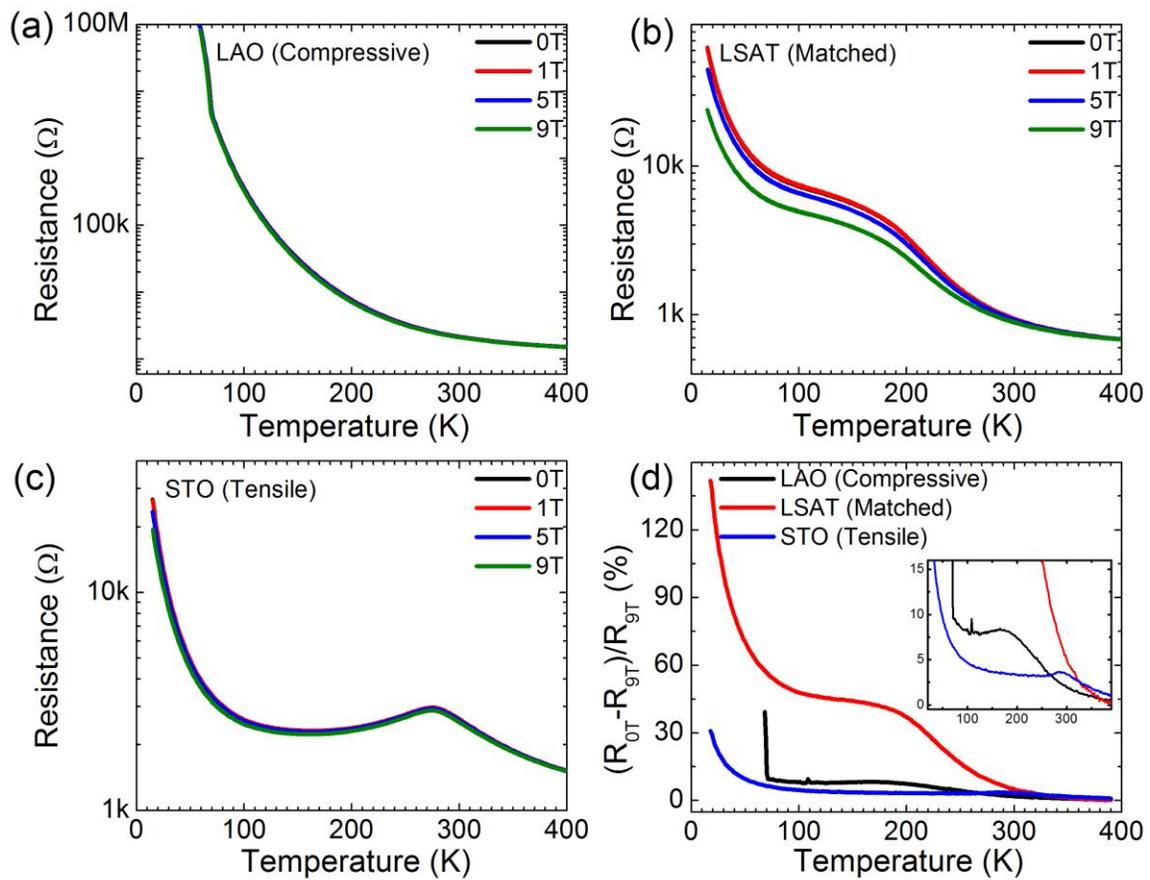

Figure 2. Resistance as a function of temperature loops under increasing out-of-plane magnetic fields for (a) compressively strained film on LAO, (b) lattice matched film on LSAT, and (c) tensile strained film on STO. (d) MR at 9T for each of the 3 films.

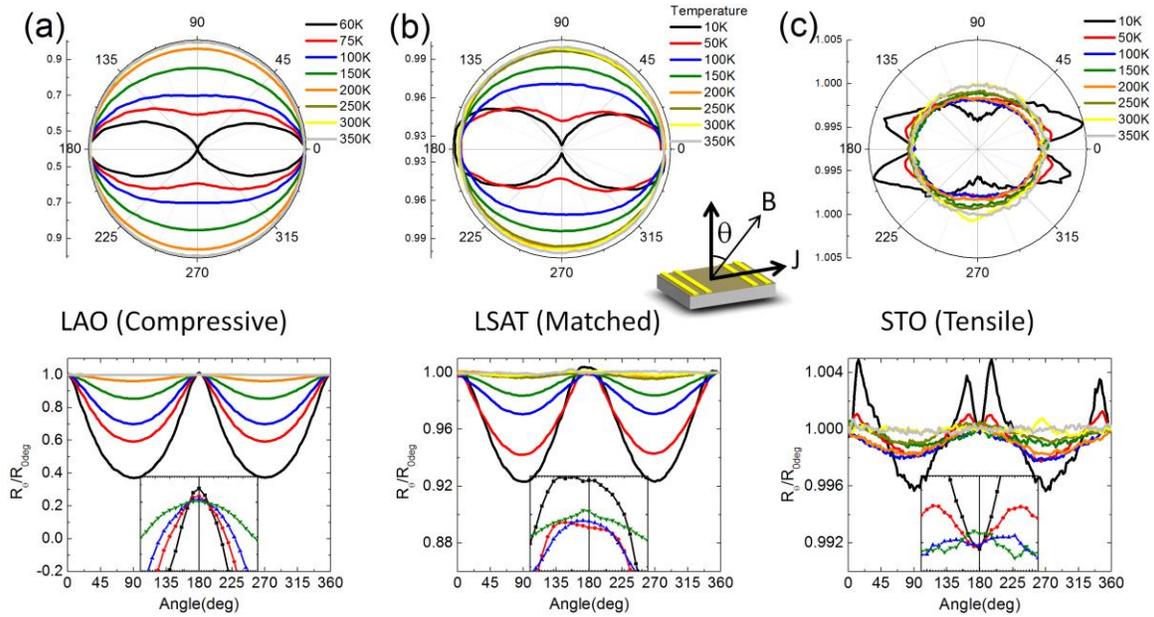

Figure 3. Angular dependent AMR with H = 9 T, the inset shows the measurement geometry where θ is the angle between H and J for (a) compressively strained film on LAO, (b) lattice matched film on LSAT, and (c) tensile strained film on STO. Insets show regions +/- 20° around field perpendicular to highlight anomalous low angle AMR behavior.

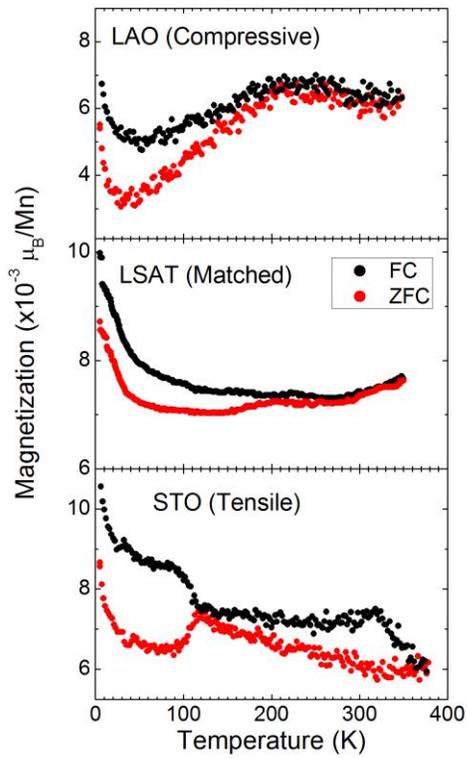

Figure 4. Field cooled (FC) and zero field cooled (ZFC) magnetization values under a 30 mT magnetic field from 5 K to 350 K show signatures of magnetic ordering.

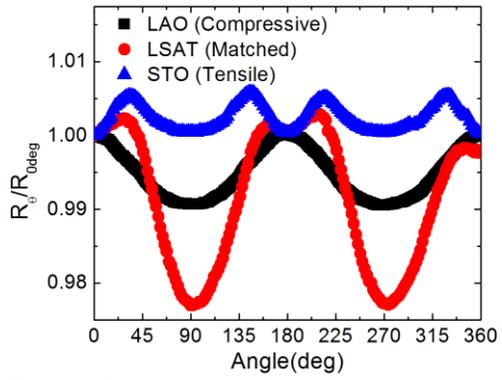

Figure 5. Low field angular dependent AMR for LAO and STO at 10 K under a 3 T magnetic field and for LAO at 60K under a 2 T magnetic field.